\documentclass[twocolumn]{aastex63}
\usepackage{amsmath}

\received{June 1, 2019}
\revised{January 10, 2019}
\accepted{\today}
\submitjournal{ApJ}

\shorttitle{COSMOS environmental effects}
\shortauthors{Shi et al.}
\begin{document}

\title{Revisiting the environmental impact on star formation activities of galaxies}

\correspondingauthor{Ke Shi}
\email{shike.astroph@gmail.com}

\author{Ke Shi}

\affiliation{School of Physical Science and Technology, Southwest University, Chongqing 400715, China}

\author{Nicola Malavasi}

\affiliation{Universitäts-Sternwarte, Fakultät für Physik, Ludwig-Maximilians-Universität, Scheinerstr. 1, 81679 München, Germany}

\author{Jun Toshikawa}

\affiliation{Nishi-Harima Astronomical Observatory, Center for Astronomy, University of Hyogo, 407-2 Nishigaichi, Sayo-cho, Sayo, Hyogo 679-5313, Japan}

\author{Xianzhong Zheng}

\affiliation{Purple Mountain Observatory, Chinese Academy of Sciences, 10 Yuan Hua Road, Nanjing  210023, China}
\affiliation{School of Astronomy and Space Sciences, University of Science and Technology of China, Hefei 230026, China}

%\author{Kyoung-Soo Lee}

%\affiliation{Department of Physics and Astronomy, Purdue University, 525 Northwestern Avenue, West Lafayette, IN 47907, USA}

\begin{abstract}
We present a systematic study of the environmental impact on star formation activities of galaxies using a mass-complete sample of $\sim$170k galaxies at $z<4$ from the latest COSMOS2020 catalog. At $z<1$, we find that the mean star-formation rate (SFR) of all galaxies decreases with increasing density of the environment. However when we consider only star-forming galaxies, the mean SFR becomes independent of the environment at $z<1$. At $z>2$ we observe a clear positive correlation between the SFR and density of the environment for all the galaxies. On the other hand, stellar mass of the galaxies increases significantly with the environments at all redshifts except for star-forming galaxies at $z<1$. The fraction of quiescent galaxies increases with increasing density of environment at $z<2$, and the ``morphology-density'' relation is confirmed to be present up to $z\sim1$. We also find that environmental quenching is negligible at $z>1$, whereas mass quenching is the dominant quenching mechanism for massive galaxies at all redshifts. Based on these results, we argue that stellar mass regulated physical processes might be the major driving force for star formation activities of galaxies. At low redshift ($z<1$) massive galaxies are quenched primarily due to their high mass, resulting in a normal ``SFR-density'' relation. At high redshift ($z>2$) most of the galaxies are star-forming ones tightly following the star-forming main sequence, and the difference in their stellar mass at different environments naturally leads to a reversal of ``SFR-density'' relation.

\end{abstract}



\section{Introduction} \label{sec:intro}
It is well known that galaxy formation and evolution is strongly affected by the local environments in which galaxies reside. In the local universe, cluster galaxies are often found to be ellipticals while late type galaxies such as spirals are mostly located in the low-density field. This is the so-called ``morphology-density'' relation which has been reported extensively in the literature \citep[e.g.,][]{Dressler80,Dressler97, Goto03, Kauffmann04}. As early type galaxies are usually quiescent and late type galaxies are more actively forming stars, a ``SFR-density'' relation can be seen as a consequence. However, it is not clear whether the ``SFR-density'' relation found in the local Universe still holds in the distant Universe. 

Observational evidence suggest that cluster galaxies formed most of their stars at early evolutionary stages and quenched rapidly afterwards \citep{Stanford98,Thomas05,Snyder12,Mart18}, implying that the ``SFR-density'' relation may be reversed at high redshift (e.g., $z>1$), such that star-formation activities are elevated in denser environments. While many studies have observed this reversal \citep[e.g.,][]{Cucciati06, Elbaz07,Cooper08,Tran10,Koyama13,Alberts14,Shimakawa18,Shi20,Lemaux22}, there are also many cases where no obvious reversal were identified \citep[e.g.,][]{Grtzbauch11,Scoville13,Darvish16,Malavasi17}. Others even found the same relation as in the local Universe \citep{Patel09,Muzzin12,Chartab20}. The discrepancy in different studies highlights the importance of a systematic study of environmental dependence on galaxy properties, especially at high redshift.

In addtion to local environment, stellar mass can also have a huge impact on star-formation activities of galaxies, such as the presence of the star-forming `main sequence' where more massive galaxies tend to have higher SFRs \citep[e.g.,][]{Daddi07,Elbaz07,Noeske07,Rodighiero11,Reddy12,Speagle14,Salmon15,Santini17}. Meanwhile, more massive galaxies are more likely to be quiescent as they consumed a significant portion of their available gas reservoirs for forming stars earlier in their history. Hence it is also important to study the effects of mass when considering the environmental effects.

While it is still under intense debate how galaxies cease their star-formation, almost all of the proposed quenching mechanisms can be categorized into two types generally known as ``environmental quenching'' and ``mass quenching'' \citep{Peng10}. The former refers to external processes occur in dense environments such as ram pressure striping \citep[e.g.][]{Gunn72}, strangulation \citep[e.g.,][]{Larson80,Balogh00} and galaxy harassment \citep[e.g.,][]{Moore96}. The latter is attributed to internal processes such as AGN and stellar feedback \citep[e.g.,][]{Croton06,Ceverino09,Fabian12,Cicone14} which usually correlate with stellar mass of galaxies. Although both mechanisms play crucial roles in quenching of galaxies, it seems that the relative importance of these mechanisms depends on the cosmic epoch and the mass range of galaxies considered \citep{Darvish16,Kawinwanichakij17,Chartab20}. Studying how mass and environmental quenching evolve across cosmic time can provide us with invaluable information on the quenching processes and help us better understand the mechanisms driving galaxy formation and evolution (``nature'' vs  ``nurture'').

The Cosmic Evolution Survey (COSMOS) is currently one of the deepest multiwavelength survey \citep{Scoville07} that covers 2 deg$^2$ of the sky. Its large contiguous area and data depth makes it a perfect laboratory to study how galaxy properties are affected by environments. \cite{Darvish16} first made use of the COSMOS UltraVISTA catalog \citep{McCracken12,Ilbert13} to investigate possible environmental trends in the COSMOS field. Using a mass complete sample of $\sim$70k galaxies, they explored in detail the role of local environment of galaxies and their stellar mass on galaxy properties out to $z\sim3$. A decade has passed since the release of the COSMOS UltraVISTA catalog, and the latest COSMOS photometric catalog, COSMOS2020 \citep{Weaver22}, marks a major improvement over the previous versions. Previous work that systematically investigated environmental effects on galaxies mostly limit their studies to $z\lesssim3$ \citep[e.g.,][]{Darvish16,Kawinwanichakij17,Chartab20}, owing to the small sample size at higher redshifts. COSMOS2020 catalog includes $\sim$1,000k galaxies, among which $>30$k galaxies are at $z>3$, making it suitable to study enviromental trends at even higher redshifts than previous studies.  On that account, it is meaningful to revisit the environmental impact on galaxy properties utilizing the COSMOS2020 catalog, which has much wider and deeper coverage along with larger data volumes than previous studies probed.

In this paper, we explore the environmental trends of galaxy properties using a mass-complete sample of galaxies at $z<4$ from the COSMOS2020 catalog. This paper is organized as follows. In Section~\ref{sec2} we describe the data selected from the COSMOS2020 catalog, as well as the method used to define the local environments of galaxies. In Section~\ref{sec3} we present the results of environmental effects on galaxy properties at different redshifts. In Section~\ref{sec4} we discuss in detail the quenching efficiency driven by stellar mass and environment. We also discuss the possible quenching mechanisms, as well as the implication of the reversal of ``SFR-density'' relation found in our study.  We summarize our results in Section~\ref{sum}. Throughout this paper we use the Planck cosmology from \cite{Planck16}. All magnitudes are given in the AB system \citep{Oke83}. Distance scales are given in comoving units unless noted otherwise.

\section{Data and Methods} \label{sec2}
In this section, we describe the data and sample selection in this work, as well as our methods of defining the environments of galaxies. 
\subsection{Data} \label{data}
This work is based on the COSMOS2020 photometric catalog \citep{Weaver22}. Comparing with the previous version \citep{Laigle16}, the new catalog includes ultra-deep optical data from Hyper Suprime-Cam (HSC) Subaru Strategic Program (SSP) PDR2 \citep[SSP;][]{Aihara19}, the deeper $u^*$ and new $u$ band imaging from the Canada France-Hawaii Telescope (CFHT) program CLAUDS \citep{Sawicki19}, the fourth UltraVISTA data release \citep{Moneti23} reaching one magnitude deeper in $K_S$ band over the entire area, and the inclusion of all \textit{Spitzer} IRAC data in the COSMOS \citep{Moneti22}.

COSMOS2020 includes roughly 966,000 sources, yielding accurate measurements of photometric redshift as well as physical properties of galaxies. The catalogs are generated using two independent photometric methods for extracting sources. \textsc{The CLASSIC} catalog performs standard aperture photometry using \textsc{SExtractor} \citep{Bertin96} on PSF-homogenized images for optical/near-infrared bands and utilizes \textsc{IRACLEAN} software \citep{Hsieh12} to perform photometry on Spitzer/IRAC images. \textsc{The FARMER} catalog, on the other hand, is produced using a new profile fitting method based on the source modelling code \textsc{The Tractor} \citep{Lang16}. Although the two methods perform equivalently well and are in high agreement with each other, \textsc{The FARMER} is able to detect fainter and higher density of high-z sources which may be due to its ability to de-blend sources at fainter magnitudes \citep{Weaver22}. In this work, we utilize \textsc{The FARMER} catalog since it provides more high-$z$ sources which would benefit our study of environmental effects at high redshifts.

\textsc{The FARMER} catalog contains a total number of 964,506 sources. In order to have a deep and clean catalog, we use the flags that mask out bright stars and edges of the HSC and Suprime-Cam images, and also require to be in the UltraVISTA regions (i.e., FLAG\_COMBINED=0 in the \textsc{The FARMER} catalog). To separate galaxies from stars/AGNs, we also use the classification in the catalog (lp\_type=0) which is derived from the SED-fitting code \textsc{LePhare} \citep{Arnouts99, Ilbert06} by combining morphological and SED criteria. This leaves us with a total number of 711,290 galaxies in the catalog.

We limit our study to galaxies at $z_\textrm{phot}<4$. This is because on one hand, the number of galaxies is too small for us to give a proper definition of the environment (for example, $<10$k galaxies are at $4<z<5$ and $<5$k galaxies are at $z>5$), considering the large field coverage of the COSMOS survey. On the other hand, the larger redshift uncertainties at higher redshifts also brings some difficulties in interpreting the results. Since the COSMOS2020 catalog used spectral energy distribution (SED) fitting method to derive the photometric redshifts with only optical-MIR data, the lack of FIR-radio information would make it challenging to capture the full spectral range of the source accurately. It is possible that a lot of low redshift galaxies could be mis-identified as high redshift sources due to confusion between Balmer break at lower redshift
and Lyman break at higher redshift, which has already been noticed in some studies \citep[e.g.,][]{Dunlop07,Shi191,Zavala23}. 

We further test the robustness of the photometric redshift in the COSMOS2020 catalog using the COSMOS Super-deblended catalog \citep{Jin18} as ancillary data. This catalog includes 194,428 galaxies in the COSMOS field with far-infrared to radio information. We cross match their catalog with ours using 1$\arcsec$ radius, finding 118,128 matches. Among them, 1205 galaxies are at $4<z<5$ in our catalog, in which we find 871 (72\%) actually have IR-radio derived photometric redshift at $z<4$. This means that majority of galaxies at $4<z<5$ (derived from optical-MIR data) could actually be dusty galaxies at lower redshift. In contrast, for galaxies at $3 < z < 4$ in the COSMOS2020 catalog, only $\sim$36\% objects are at $z < 3$ in the Super-deblended catalog; for galaxies at $2 < z < 3$ in the COSMOS2020 catalog,
only $\sim$21\% sources have IR-radio photometric redshift at $z < 2$. Thus to eliminate low-z contaminations as much as possible, we confine our study to galaxies at $z<4$ in the COSMOS2020 catalog.

Lastly, recent observations have revealed a population of massive galaxies at $z>3$, which are too faint to be detected in rest-frame UV-optical \citep{Wang19} or NIR \citep{Yamaguchi19}. These massive galaxies are proved to be heavily dust-obscured star-forming galaxies that are detected in the Atacama Large Millimeter/submillimeter Array (ALMA), as well as in the JWST NIRCam imaging \citep[e.g.,][]{Barrufet23,Nelson23}. In this sense, the COSMOS2020 catalog (based on UV-NIR data) could be highly incomplete at high-mass end beyond $z>3$, therefore we caution the study of environmental effects on $z>3$ galaxies although we still present their results in the following sections for a general comparison.

COSMOS2020 catalog also classifies galaxies into two different types: star-forming and quiescent. Following \cite{Ilbert13}, quiescent galaxies are defined to have rest-frame colors $M_\textrm{NUV}-M_r>3(M_r-M_J)+1$ and $M_\textrm{NUV}-M_r>3.1$ in the NUV$-r$ vs. $r-J$ diagram. The stellar-mass completeness of the COSMOS2020 sample is calculated in \citet{Weaver22} with masses reported by \textsc{LePhare} considering magnitude limits of IRAC channel 1, based on the method described in \cite{Pozzetti10}. The mass completeness is determined for both star-forming galaxies and quiescent galaxies shown in Figure 20 in \cite{Weaver22}. Since the mass completeness limit increases with increasing redshift due to the selection effect, higher redshift samples would be strongly biased towards more massive galaxies compared to those at lower redshifts. The goal of this paper is to compare the environmental dependence of galaxy properties at different redshifts, accordingly it is important that we select galaxies in an unbiased way to further define the environments and to study galaxy properties. To this end, we adopt the mass completeness of the highest redshift bin studied in this work, which is $10^{9.26}M_\odot$ for star-forming galaxies and $10^{9.64}M_\odot$ for quiescent galaxies at $z\approx4$. This represents the low-mass limit of our sample.

Our final sample includes 173,339 galaxies, among which 17,045 are quiescent. We divide the whole sample into six redshift bins, and each bin contains at least $\sim$20k galaxies to ensure our results are not biased by small number statistics. The sample size of each bin is summarized in Table \ref{table1}. The physical properties such as stellar masses, SFRs are derived via SED-fitting technique (optical-MIR) using \textsc{LePhare} in the COSMOS2020 catalog. We refer the readers to \cite{Weaver22} for a detailed description of the methods. Last but not least, since the SFR estimates used here is based purely on optical-MIR data, it may not be that accurate in lack of FIR-submm information. In fact, many studies have revealed the prevalence of dusty star-forming galaxies at high redshift, especially in high-density environments \citep[e.g.,][]{Casey16, Kato16,Umehata17, Harikane19, Zhang22}. These massive dusty galaxies usually have extremely high SFRs and are generally invisible in rest-frame UV-NIR due to heavily dust obscuration. Thus for these galaxies the SFRs in this work might be significantly underestimated. Future work using FIR-submm data could give a more accurate estimate of the SFRs of these dusty star-forming galaxies (Jin et al. in preparation). Bearing that in mind, in the following sections, we explore the environmental impact on the physical properties of galaxies in detail. 

\subsection{Galaxy Environment Measurement}

Galaxy environment is defined as the local density field in which that galaxy resides. In order to construct the density field of galaxies, one needs accurate redshift measurements first. Thanks to the state-of-art COSMOS2020 catalog, we now have highly accurate photometric redshifts (photo-z) for several hundred thousands of galaxies that are selected in a consistent way. The photometric redshifts are derived using \textsc{Lephare}. The photo-z uncertainty is $\sim0.01-0.02$ at $z<1$, increasing to $\sim0.1$ up to $z\sim4$ in \textsc{The FARMER} catalog. Taking the uncertainties of photo-z into account, we construct a series of narrow redshift slices and regard each slice as a 2D structure. The width of each slice is defined to be the median value of photo-z uncertainty $\Delta z=z_\textrm{max}-z_\textrm{min}$ at that redshift, where $z_\textrm{max}$ ($z_\textrm{min}$) is the upper (lower) limit (68\% confidence level) of the best-fit photometric redshift. In the end, 81 slices are selected spanning from $z=0.1$ to $z=4$. We also note that the results of this paper are insensitive to the redshift slice we choose. In fact we have tried to set the redshift slice to a constant comoving size of 40 Mpc (e.g., $\Delta z\sim0.01$ at $z=0.1$ and $\Delta z\sim0.05$ at $z\sim4$),  and our main results in the following sections remain unchanged.

After deviding the whole redshift space into a number of redshift slices, we now have obtained a series of two-dimensional planes where we can perform density estimate for the galaxies in those planes. There are many ways to measure galaxy densities, such as using a fixed kernel \citep[e.g.,][]{Hayashino04,Lee14,Malavasi16,Harikane19,Shi191} or weighted adaptive kernels \citep[e.g.,][]{Yang10,Darvish15,Chartab20}, or certain scale independent methods such as Voronoi tessellation \citep[e.g.,][]{Darvish15,Dey16,Shi21} or Delaunay tessellation \citep[e.g.,][]{Marinoni02,Darvish15,Malvasi21}. \cite{Darvish15} tested the performance of different density estimators in the COSMOS field and found that the Voronoi tessellation and adaptive kernel smoothing outperform other methods. In this work, we use Voronoi tessellation to estimate the 2D density of galaxies. Voronoi tessellation is a non-parametric density estimator that has been widely used in astronomy field to detect both clusters \citep[e.g.,][]{Ramella01,Soares11,Euclid19} and protoclusters \citep[e.g.,][]{Dey16,Shi21,Brinch23}, which we briefly describe below.

A Voronoi tessellation is a unique way of dividing a 2D distribution of galaxies into convex cells, with each cell containing only one galaxy. The vertices of each cell are closer to that enclosed galaxy than any other in the entire plane. Based on this definition, the denser the regions, the smaller the cell area. As a consequence, the local density ($f_i$) of each cell can be measured as the inverse of the cell area ($A_i$):

\begin{equation}
f_i=1/A_i.
\end{equation}
To determine the relative density contrast of each cell, we calculate the average density ($\hat{f_i}$) of all the cells in the entire plane, and the density contrast ($\Sigma$) of each cell is then $\Sigma=f_i/\hat{f_i}$. Note that some cells in the boarder area might be open, hence we exclude these cells in our calculation of the density values. 

\begin{figure}[ht!]
\epsscale{1.2}
\plotone{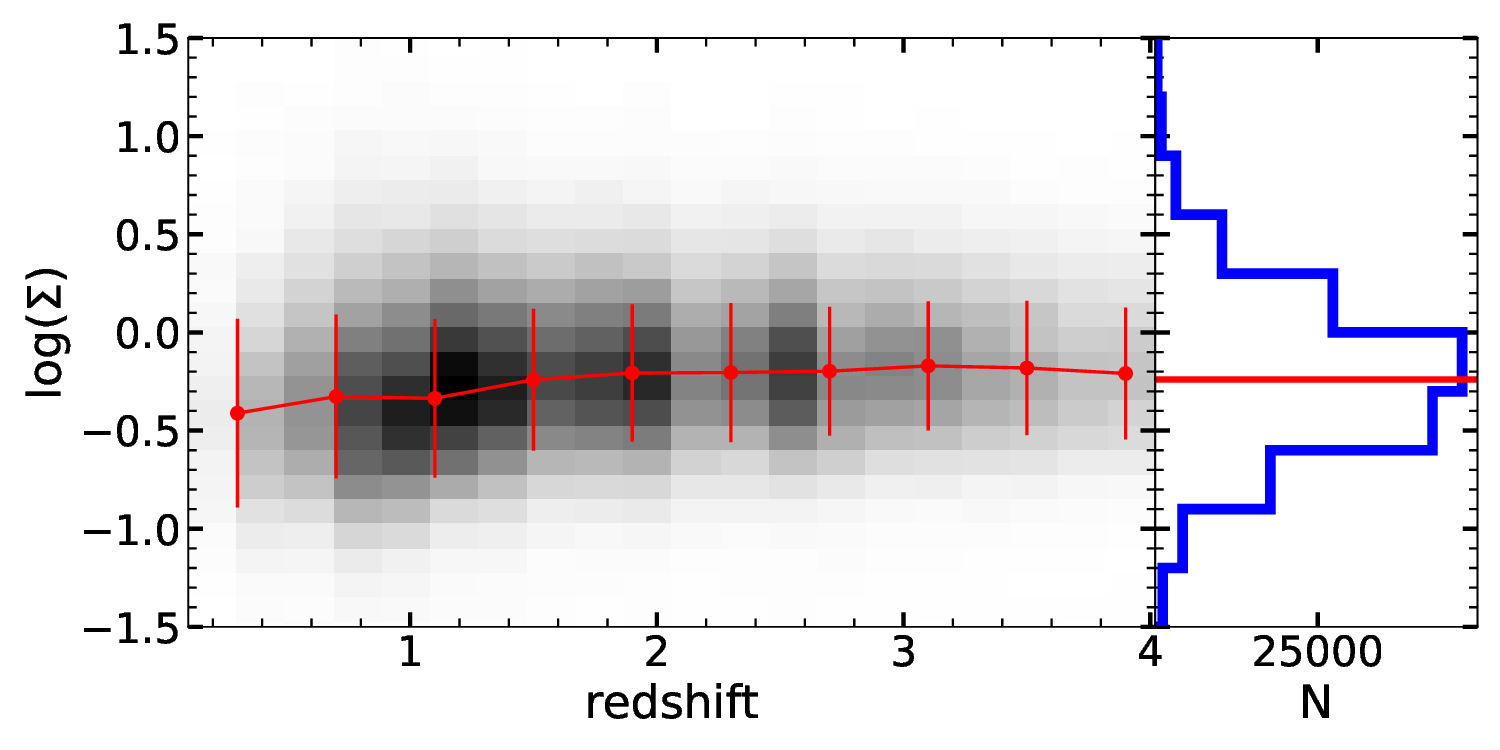}
\caption{
  Density contrast as a function of redshift. The shaded areas show the distribution of galaxies. The median values of density contrast in each redshift bin are shown with errorbars denoting standard deviations. The histogram of density contrast is also shown where the red line denoting the median value of all galaxies.
}
\label{redshift}
\end{figure}

Figure \ref{redshift} shows the evolution of density contrast $\Sigma$ as a function of redshift. We see that $\Sigma$ is almost independent of redshift, thus we are confident that our study of environmental effects in the following sections is not biased by the redshift evolution of density measurements. It is also noted that the median density contrast is smaller than 1. This is probably due to the fact that very high-density regions such as clusters strongly elevate the average density values. Since most galaxies are located in lower-density regions such as groups rather than clusters, their density values turn out to be smaller than the average.

\begin{deluxetable*}{ccc}
\tablecaption{Properties of the mass complete sample \label{table1}}
\tablehead{
\colhead{Redshift range}  & \colhead{Number of galaxies} & \colhead{Number of quiescent galaxies}
}
\startdata
$0.1\leq z<1.0$  & 36269 & 7328\\
$1.0\leq z<1.5$  & 34538 & 4435\\
$1.5\leq z<2.0$  & 26272 & 2386\\
$2.0\leq z<2.5$  & 19433 & 857\\
$2.5\leq z<3.0$  & 19920 & 584\\
$3.0\leq z<4.0$  & 22736 & 510\\
\enddata
\end{deluxetable*}

\section{Results} \label{sec3}
Using the Voronoi tessellation technique described in the above section, we obtained the density field in each redshift slice. In this section, we study the environmental effects on galaxy properties as a function of redshift.

\subsection{The Role of Environments on SFR, Stellar mass, and sSFR} \label{sec31}
\begin{figure*}[ht!]
\epsscale{1.0}
\plottwo{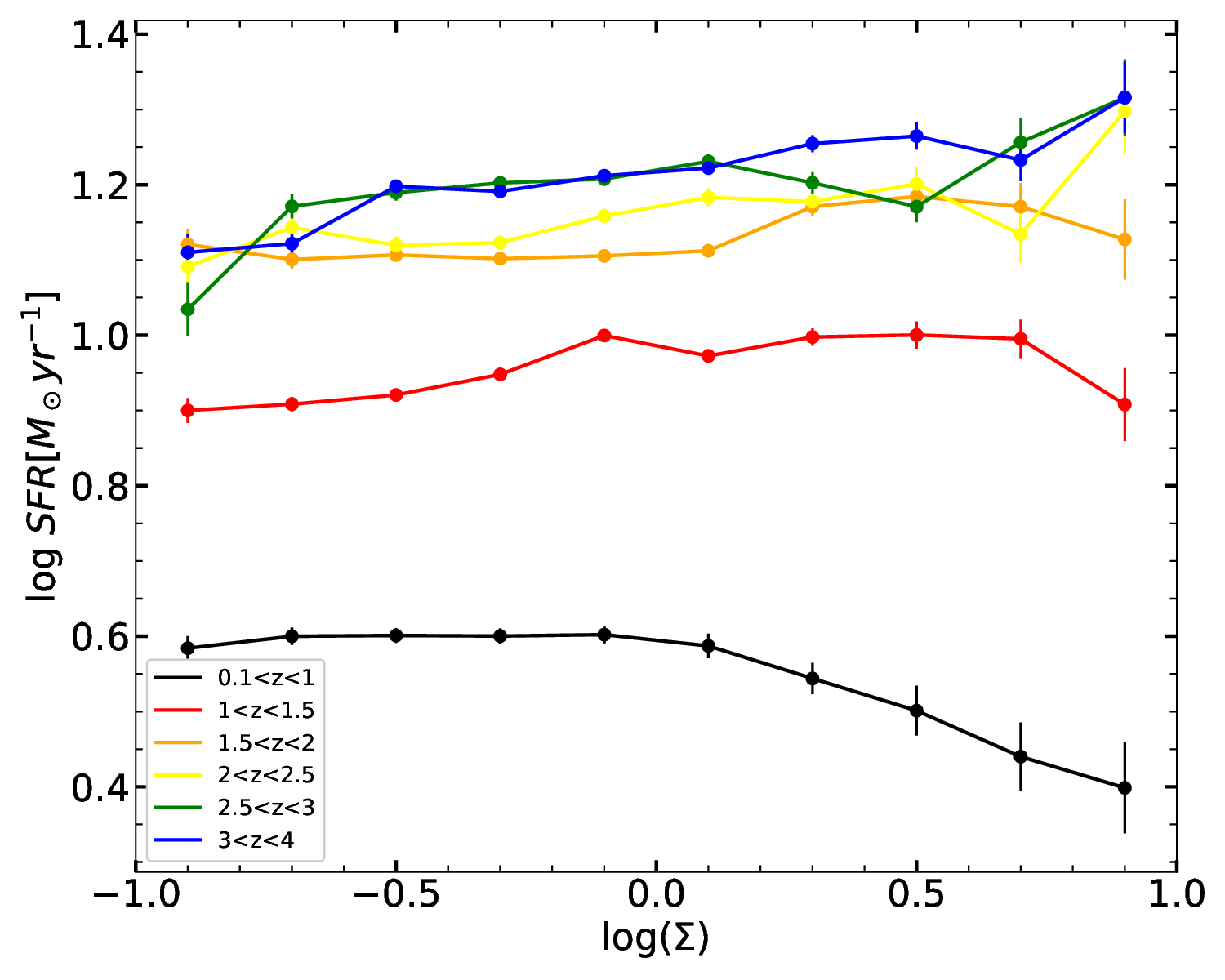}{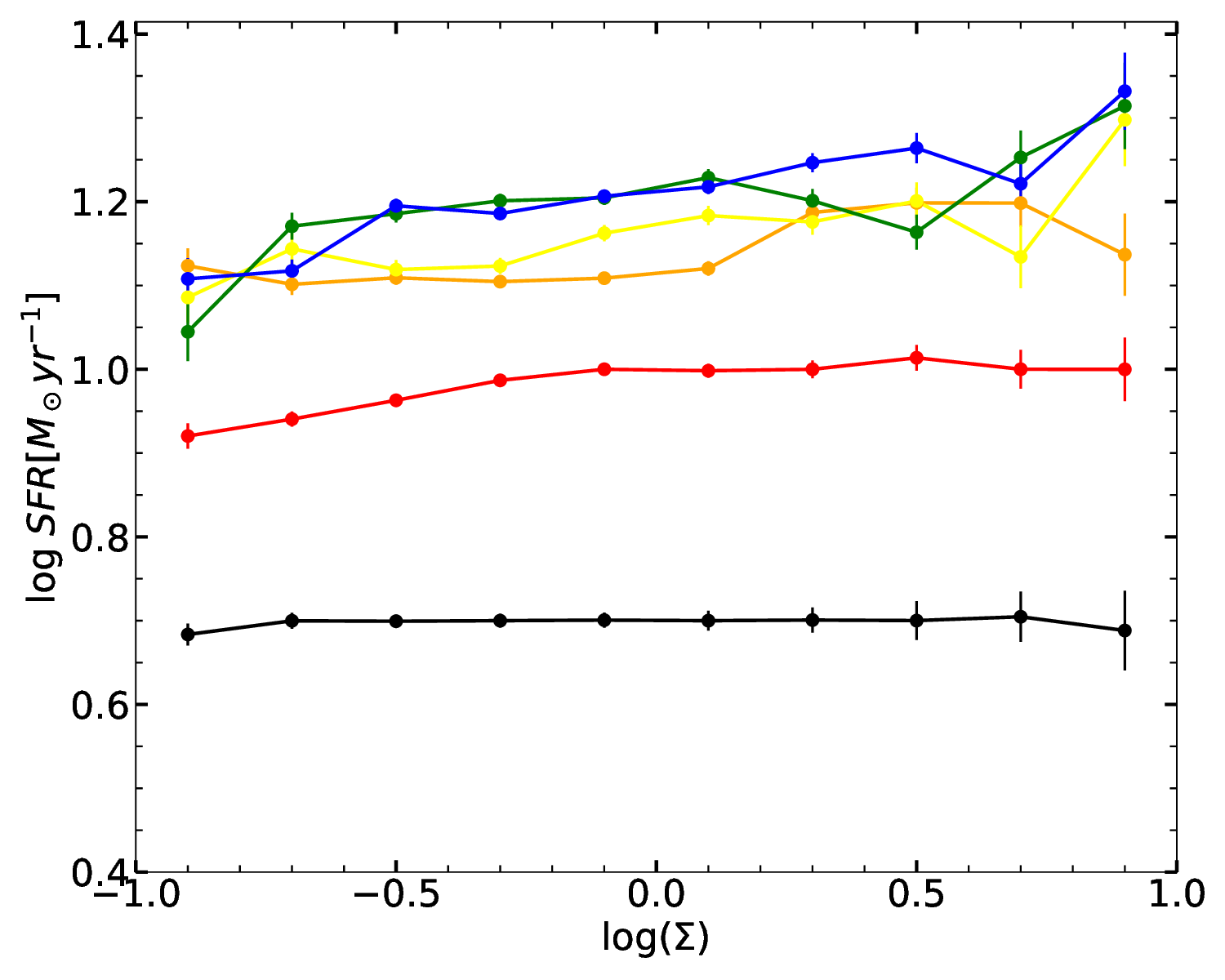}
\caption{
 {\it Left:} SFR as a function of density contrast at different redshift bins. In each density bin, the mean value of SFRs (after 3-$\sigma$ clipping) is shown with errorbar denoting the standard deviation of the mean. 
 {\it Right:} Same as the left, but for star-forming galaxies only.
}
\label{figure1}
\end{figure*}

The left panel of Figure \ref{figure1} shows the mean SFR of all the galaxies in our sample as a function of the environments at different redshift bins.  It is clear that in our lowest redshift bin $0.1<z<1$, the mean SFR decreases with increasing density of the environments. This trend disappears as we go into higher redshift to $1<z<1.5$. While at $z>2$, there appears to be a positive correlation between the SFRs and the environments. To show possible correlation in a clearer way, we resort to Kendall's tau coefficient \citep{Kendall38}, a non-parametric way to measure the strength of association between two variables\footnote{Our results do not change if instead we use Spearman's rank correlation coefficient.}. Table \ref{table2} shows Kendall's tau values for SFR and density contrast in each redshift bin. At $z<1$, a significant negative correlation can be seen ($p$-value$=$0.002). There are no obvious correlations at $1<z<1.5$ and $1.5<z<2$ ($p$-value$>$0.05), but strong positive correlations are identified at higher redshift $z>2$ ($p$-values$<$0.05). It is without doubt that we witness a reversal of `SFR-density' relation at high redshift ($z>2$). 

Since our sample includes both quiescent and star-forming galaxies, and it has been shown that different galaxy types could lead to different `SFR-density' correlation at different redshifts \citep{Darvish16}. Therefore on the right panel of Figure \ref{figure1} we show the mean SFR as a function of the environment for star-forming galaxies only. Most notably, at $z<1$, the SFR tends to be independent of the environment for star-forming galaxies ($p$-value$=$0.1). At $1<z<1.5$ the SFR begins to increase with the density of environment ($p$-value$=$0.005). While at $z>1.5$ there is no obvious difference in the `SFR-density' relation between all galaxies and star-forming galaxies. As we see in Table \ref{table1}, the fraction of quiescent galaxies drops dramatically beyond $z>1$, therefore our sample at high redshift is dominated by star-forming galaxy population, whose SFRs increase with increasing density of the environments. On the other hand, the anti-correlation at $z<1$ for all the galaxies is attributed to the enhancement of quiescent fraction at low redshift, which increases with increasing density of the environment. We will return to this topic in Section 3.2.

\begin{figure*}[ht!]
\epsscale{1.0}
\plottwo{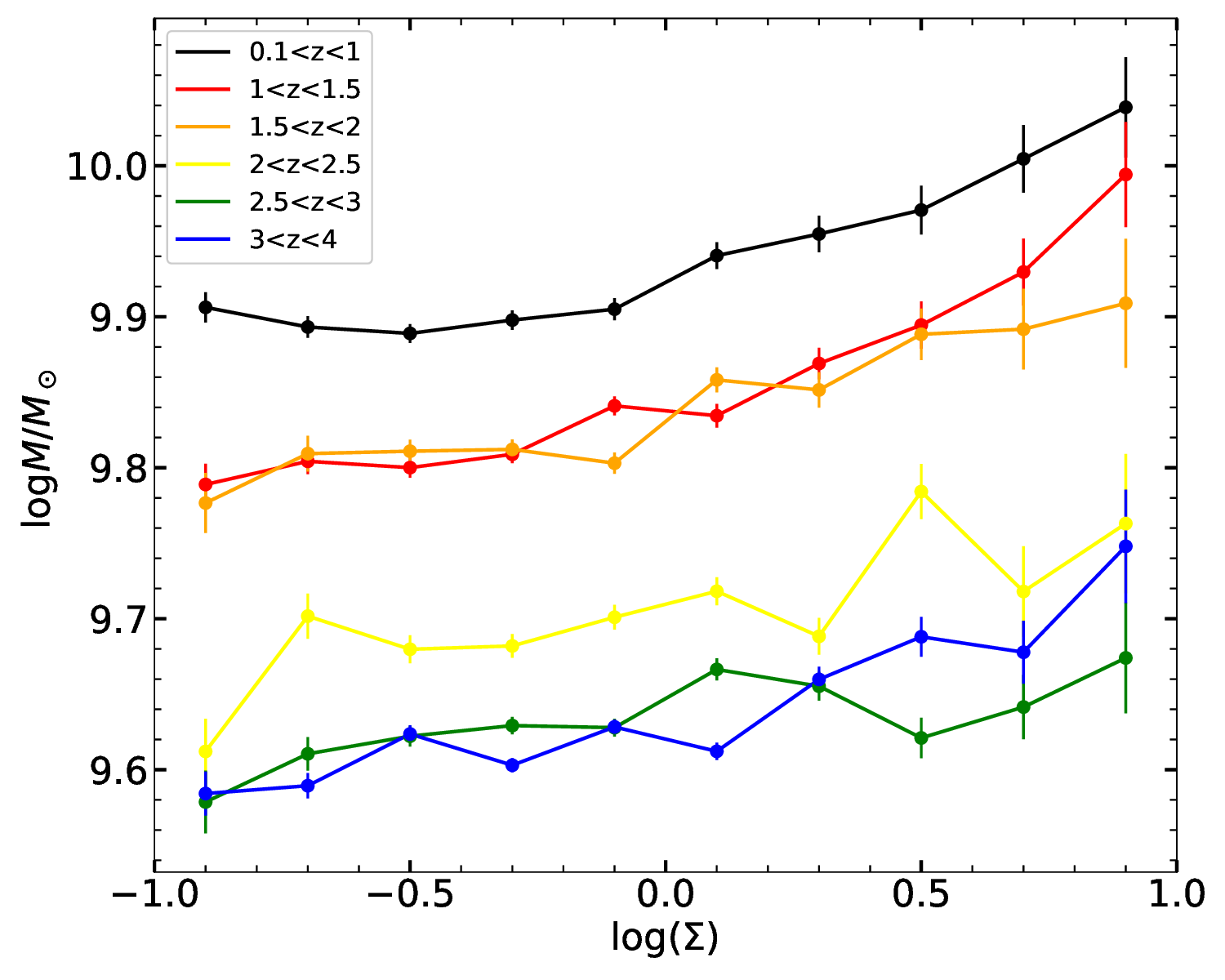}{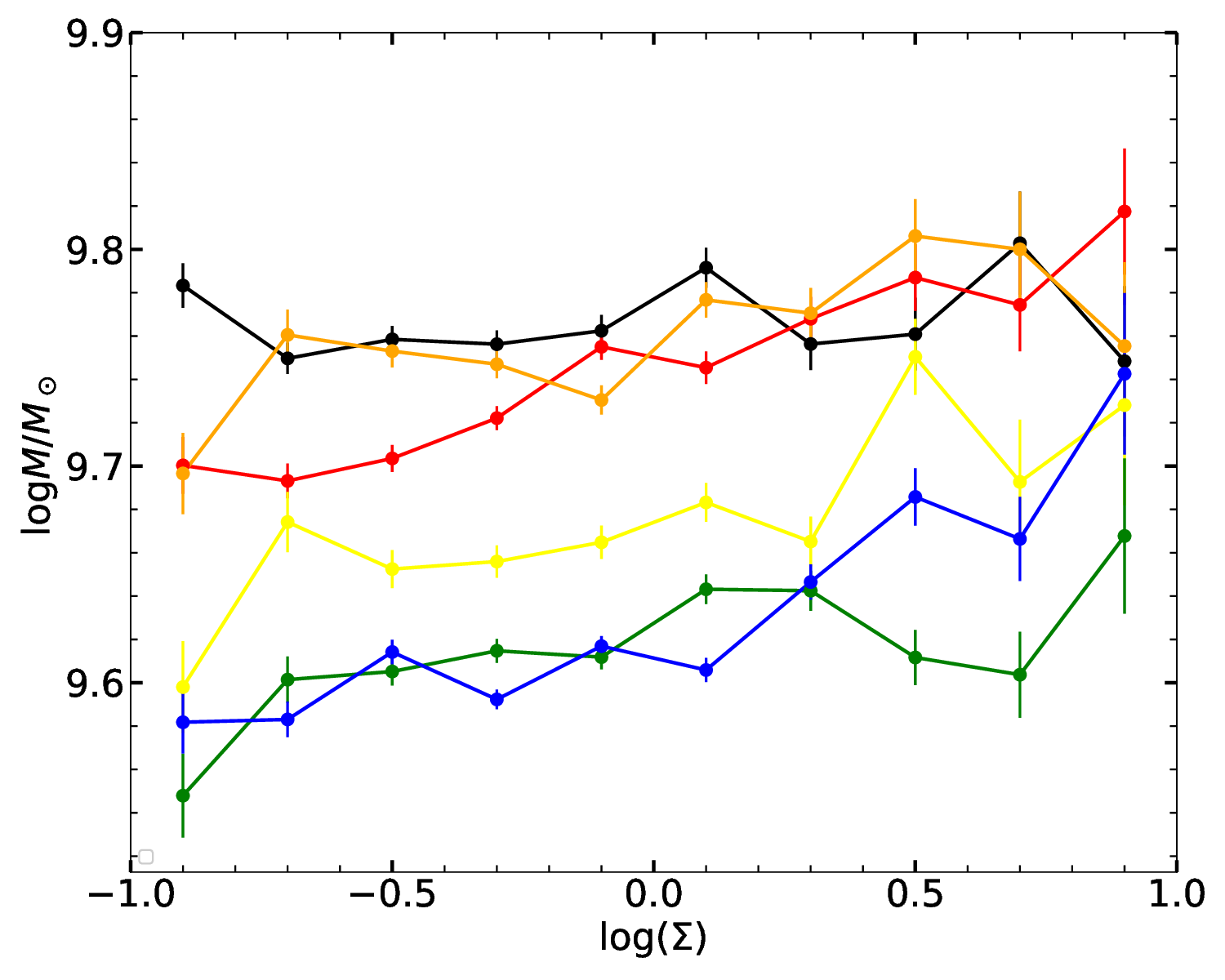}
\caption{
 {\it Left:} Stellar mass as a function of density contrast at different redshift bins. In each density bin, the mean value of stellar mass (after 3-$\sigma$ clipping) is shown with errorbar denoting the standard deviation of the mean. 
 {\it Right:} Same as the left, but for star-forming galaxies only.
}
\label{figure2}
\end{figure*}

In hierarchical structure formation, dense environments provide favorable conditions for the growth of massive galaxies via various physical processes such as galaxy mergers and accretions \citep[e.g.,][]{Kauffmann93,Kauffmann04,Weinmann06,Etherington17}. Therefore in principle stellar mass is also correlated with the density of the environment. We investigate the environmental dependence of stellar mass in Figure \ref{figure2}. The left panel shows the relation for all the galaxies while the right panel shows star-forming galaxies only. In both cases, we see a nearly monotonic increase in stellar mass with density contrast (see also Table \ref{table2}). The only exception is star-forming galaxies at $z<1$, where there appears to be no obvious trend ($p$-value$=$0.86). This discrepancy suggests that denser environments usually host more massive galaxies, which tend to be quiescent at $z<1$. On the other hand, at $z>1$, most of the galaxies are star-forming ones while they still follow the same positive trend, suggesting that they do not suffer much from the quenching mechanisms as those at $z<1$. 

\begin{figure*}[ht!]
\epsscale{1.0}
\plottwo{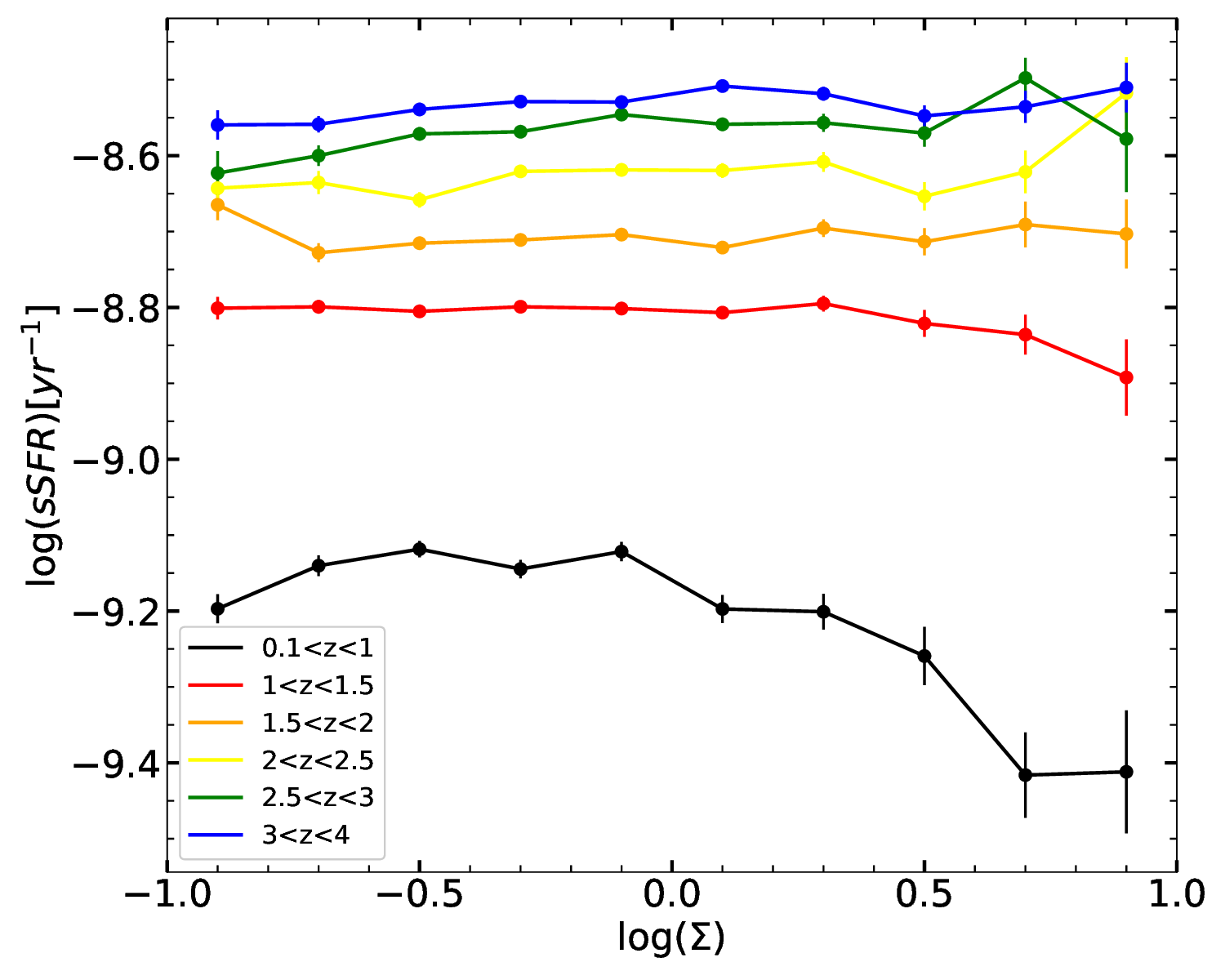}{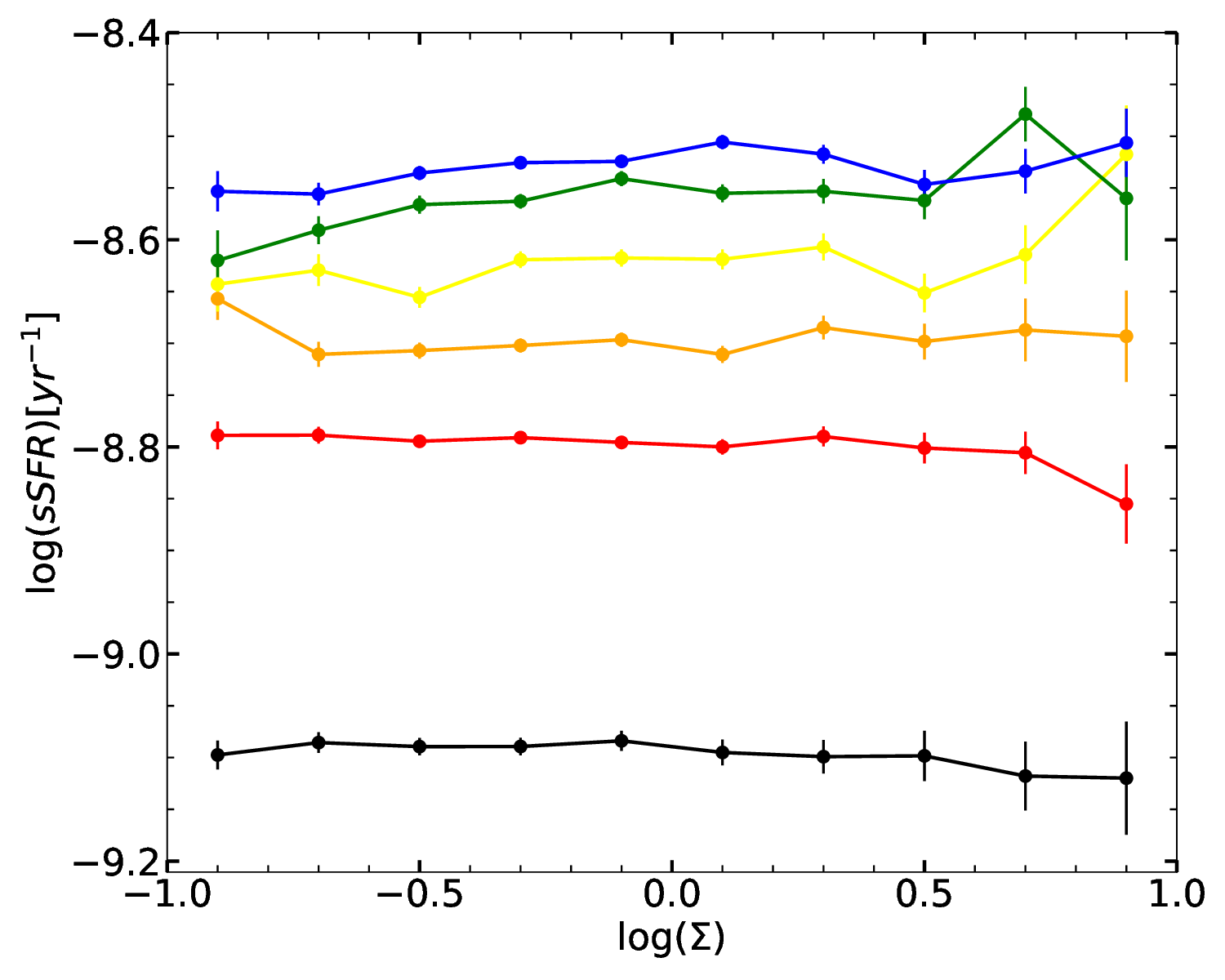}
\caption{
 {\it Left:} sSFR as a function of density contrast at different redshift bins. In each density bin, the mean value of sSFRs (after 3-$\sigma$ clipping) is shown with errorbar denoting the standard deviation of the mean.
 {\it Right:} Same as the left, but for star-forming galaxies only.
}
\label{figure3}
\end{figure*}

The SFRs of galaxies are also known to be tightly correlated with their stellar masses, resulting in a star-forming `main sequence' reported previously in the literature \citep[e.g.,][]{Daddi07,Elbaz07,Noeske07,Rodighiero11,Reddy12,Speagle14,Salmon15,Santini17}. As a consequence, the trends of SFRs we see in Figure \ref{figure1} could actually be due to the change of stellar mass in different environments. We further consider the environmental dependence of specific star-formation rate sSFR (i.e., SFR/$M$). Figure \ref{figure3} shows the mean sSFR as a function of density contrast. For all the galaxies (left panel),  a clear anti-correlation can be seen at $z<1$, which is further confirmed by Kendall's tau test in Table \ref{table2}. The trend is modest at $1<z<1.5$ ($p$-value$=$0.05). At $z>1.5$ sSFRs become independent of the environment, indicating a tight `main-sequence' at high redshift regardless of the environment \citep{Peng10,Koyama13,Darvish14}. As for star-forming galaxies (right panel), there is still a mild anti-correlation between the sSFR and the density contrast at $z<1$ ($p$-value$=$0.03) but not as significant as all the galaxies in the left panel. While at $z>1$ the trend for star-forming galaxies is almost identical to all the galaxies. These results indicate that quiescent galaxies are primarily responsible for the anti-correlation between sSFR and density of the environment at $z<1$, that denser environments tend to host more passive galaxies at low redshift. On the other hand, at $z>1.5$ galaxies are dominated by star-forming populations that follow the star-forming `main sequence' independent of the environment.

\begin{figure*}[ht!]
\epsscale{1.0}
\plottwo{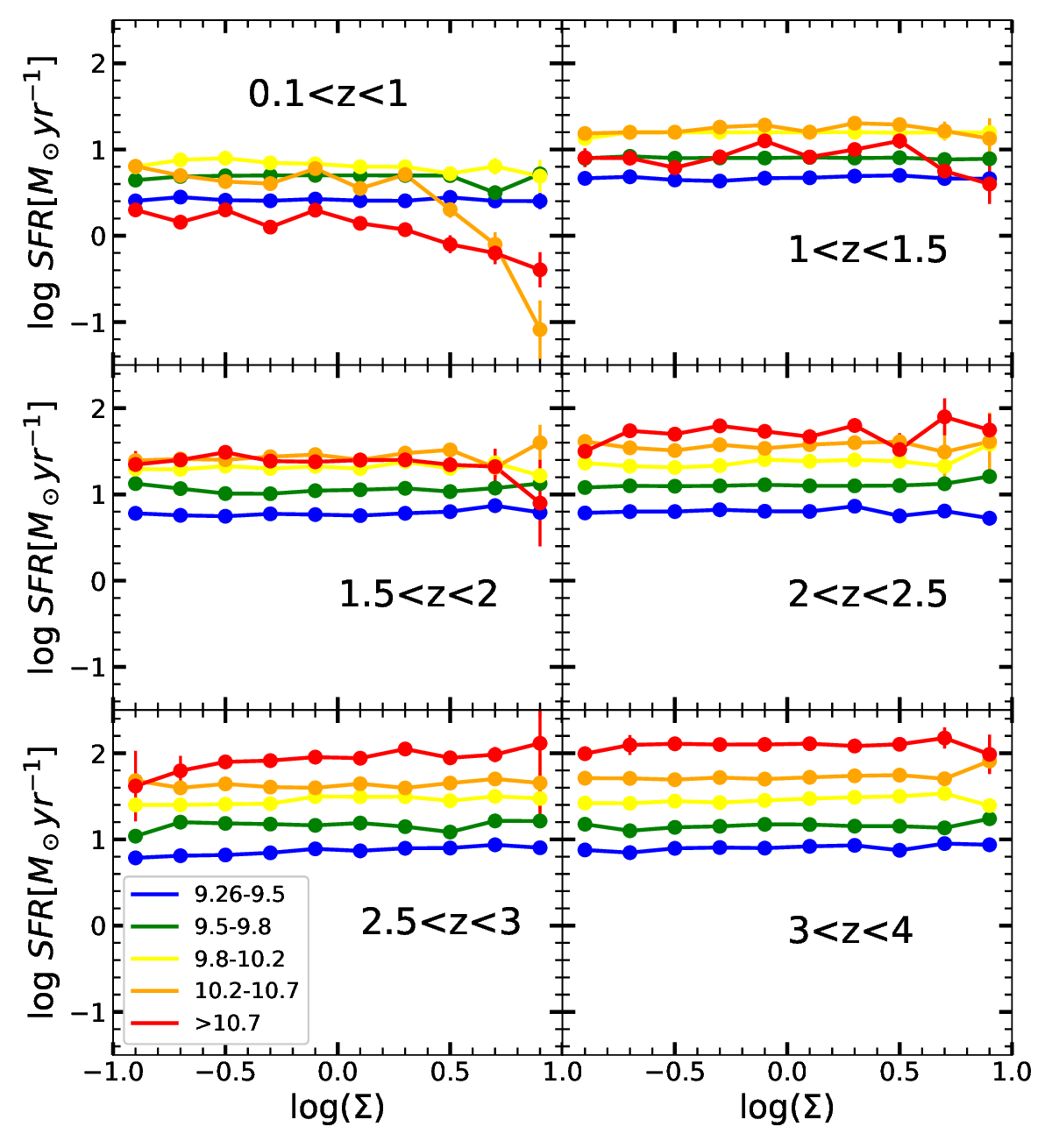}{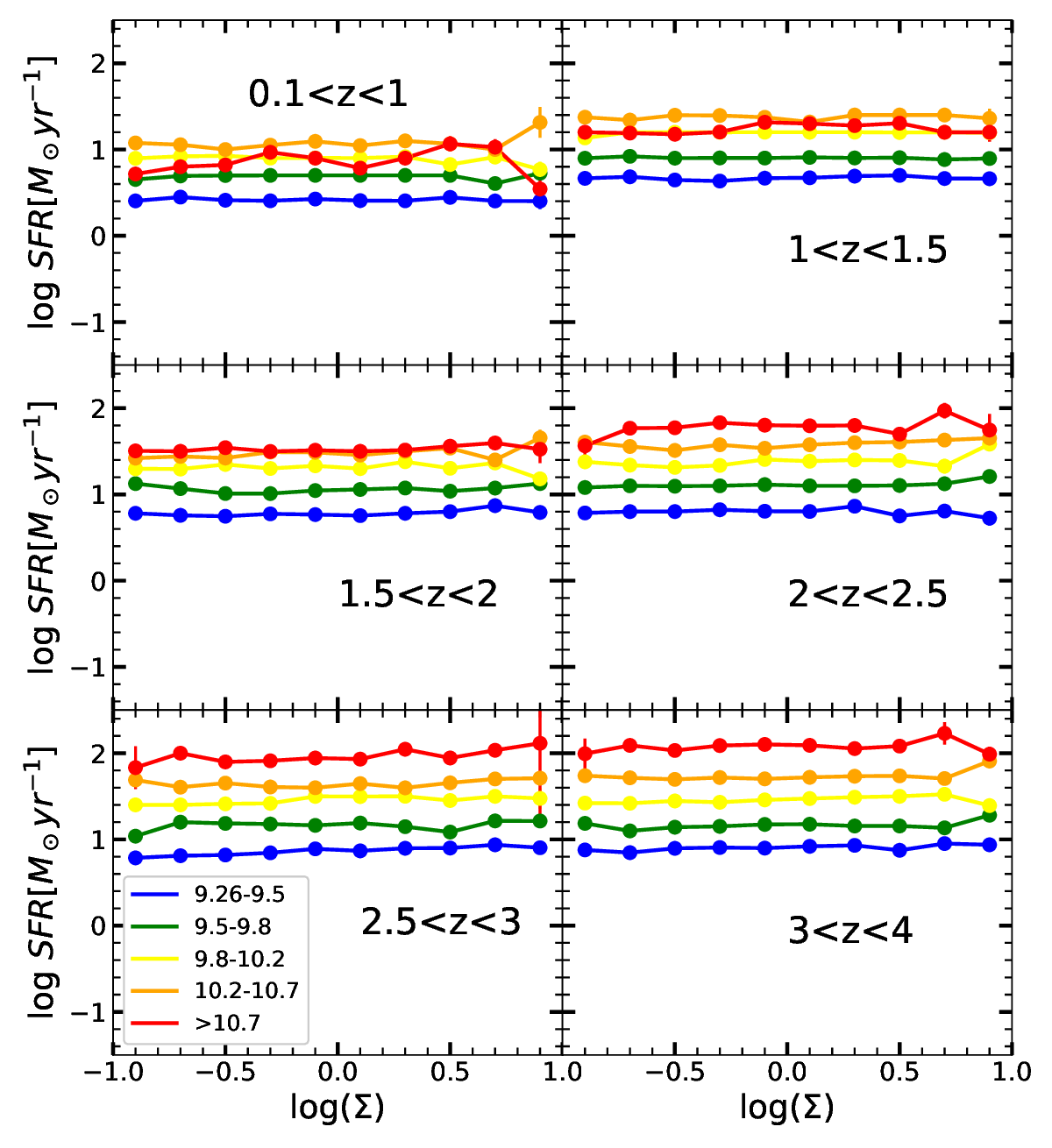}
\caption{
  {\it Left:} SFR as a function of density contrast at different redshifts for different mass bins. Different colors denote different mass bins (in log scale). In each density bin, the mean value of SFRs (after 3-$\sigma$ clipping) is shown with errorbar denoting the standard deviation of the mean. {\it Right:} Same as the left but for star-forming galaxies only.
}
\label{figure4}
\end{figure*}

To further separate the effects of mass on star formation activities of galaxies, in Figure \ref{figure4} (left panel) we investigate the environmental trends of SFRs at fixed mass bins. Interestingly, we find that at different redshifts, the trends of SFRs are quite different for different mass bins. At $0.1<z<1$, galaxies in the two highest mass bins ($M>10^{10.7}M_\odot$ and $10^{10.2}M_\odot<M<10^{10.7}M_\odot$) tend to have lower SFRs than those in lower mass bins, contrary to the star-forming `main sequence' seen in numerous studies. At $M<10^{10.2}M_\odot$, the normal `main sequence' reappears where more massive galaxies have higher SFRs.  These clearly indicate that galaxies with high stellar mass (e.g., $M>10^{10.2}M_\odot$) are more quenched due to their internal processes (i.e., mass) even in the same environments, compared to their low mass counterparts. In addition, there are clear trends that the mean SFR decreases with increasing density of the environments for the two highest mass bins, implying the role of the environment in quenching massive galaxies. At $1<z<1.5$ and $1.5<z<2$, only galaxies in the highest mass bin have suppressed star-formation activities compared to lower mass ones. As we go for higher redshifts ($z>2$), the `main sequence' is seen once again, and the SFRs tend to be independent of the environments at fixed stellar mass. If we consider only star-forming galaxies (right panel), only galaxies in the highest mass bin ($M>10^{10.7}M_\odot$) are quenched up to $z\sim2$, though their level of quenching is much smaller than that in the left panel.

Combining the results from Figure \ref{figure1}-\ref{figure4}, some insights can be gained regarding the environmental and mass dependence of star-formation activities. At $z>2$ our sample is mainly composed of star-forming galaxies, for these galaxies, the higher stellar mass in denser environment (Figure \ref{figure2}) leads to an average higher SFR seen in Figure \ref{figure4}, which is reflected as the reversal of `SFR-density' relation observed in Figure \ref{figure1}, possibly due to larger gas reservoirs or higher star-forming efficiency they have \citep[e.g.,][]{Scoville16, Wang18, Scoville23}. Meanwhile at fixed stellar mass, SFRs are independent of the environments (Figure \ref{figure4}), leading to a flat sSFR-density relation in Figure \ref{figure3}. These results indicate that environmental quenching is negligible at $z>2$.  However, at $z<1$, it is noted in Figure \ref{figure4} that massive galaxies ($M>10^{10.2}M_\odot$) have already truncated their star formation compared to less massive ones even in the same environment, a clear signal of mass quenching. Again, as these massive galaxies tend to be located in high-density regions, we see the normal `SFR-density' relation in Figure \ref{figure1}, which further leads to a clear anti-correlation between sSFR and environment in Figure \ref{figure3}. Meanwhile the SFRs of these massive galaxies decrease with increasing density of the environments, showcasing the effect of environmental quenching. At moderate redshifts ($1<z<2$), only the most massive galaxies ($M>10^{10.7}M_\odot$) have truncated their star formation (Figure \ref{figure4}), thus it may mark the transition epoch when the reversal of `SFR-density' is happening, as we see in Figure \ref{figure1} where no clear trends are observed.

Based on the above arguments, we infer that stellar mass might be the major driving force for star-formation activities of galaxies while environments may merely act as a secondary role. We will further discuss in detail both mass and environmental quenching in Section \ref{sec4}.

\begin{deluxetable*}{cccccc}[h]
\tablecaption{Kendall's tau values for all the galaxies at different redshifts \label{table2}}
\tablehead{
\colhead{Redshift range} & \colhead{SFR} & \colhead{Stellar Mass} & \colhead{sSFR} & \colhead{Quiescent fraction} & \colhead{ETG fraction}
}
\startdata
$0.1\leq z<1.0$ & -0.56 (0.03) & 0.78 (0.001) & -0.64 (0.01) & 0.91 (3$\times10^{-5}$) & 0.96 (5$\times10^{-6}$) \\
$1.0\leq z<1.5$ & 0.42 (0.10) & 0.91 (3$\times10^{-5}$) & -0.51 (0.05) & 0.73 (0.002) & 0.38 (0.16)\\
$1.5\leq z<2.0$ & 0.47 (0.07) & 0.82 (4$\times10^{-4}$) & 0.24 (0.38) & 0.82 (4$\times10^{-4}$) & -0.57 (0.06) \\
$2.0\leq z<2.5$ & 0.64 (0.01) & 0.60 (0.02) & 0.42 (0.11) & 0.46 (0.07) & -0.28 (0.29) \\
$2.5\leq z<3.0$ & 0.64 (0.01) & 0.60 (0.02) & 0.42 (0.11) & 0.11 (0.72) & -0.57 (0.06) \\
$3.0\leq z<4.0$ & 0.86 (1$\times10^{-4}$) & 0.82 (4$\times10^{-4}$) & 0.47 (0.07) & 0.07 (0.86) & -0.57 (0.06) \\
\enddata
\tablecomments{Kendall's tau values are given for all the galaxies in each redshift bin for SFR, stellar mass, sSFR, Quiescent fraction and ETG fraction respectively, with corresponding $p$-values in parentheses. Values close to 1 indicate strong concordance (i.e., positive correlation) while values close to -1 indicate strong discordance (i.e., negative correlation). Small $p$-values  (e.g., $<0.05$) indicate significant association between two variables.}
\end{deluxetable*}

\subsection{The Role of Environments on Quiescent Fraction and Morphology of Galaxies} \label{sec32}

To further investigate the role of environments on galaxy quenching, in this section we study the quiescent fraction of galaxies as a function of the environment, as well as the environmental impact on the morphologies of galaxies.

\begin{figure*}[ht!]
\epsscale{1.0}
\plottwo{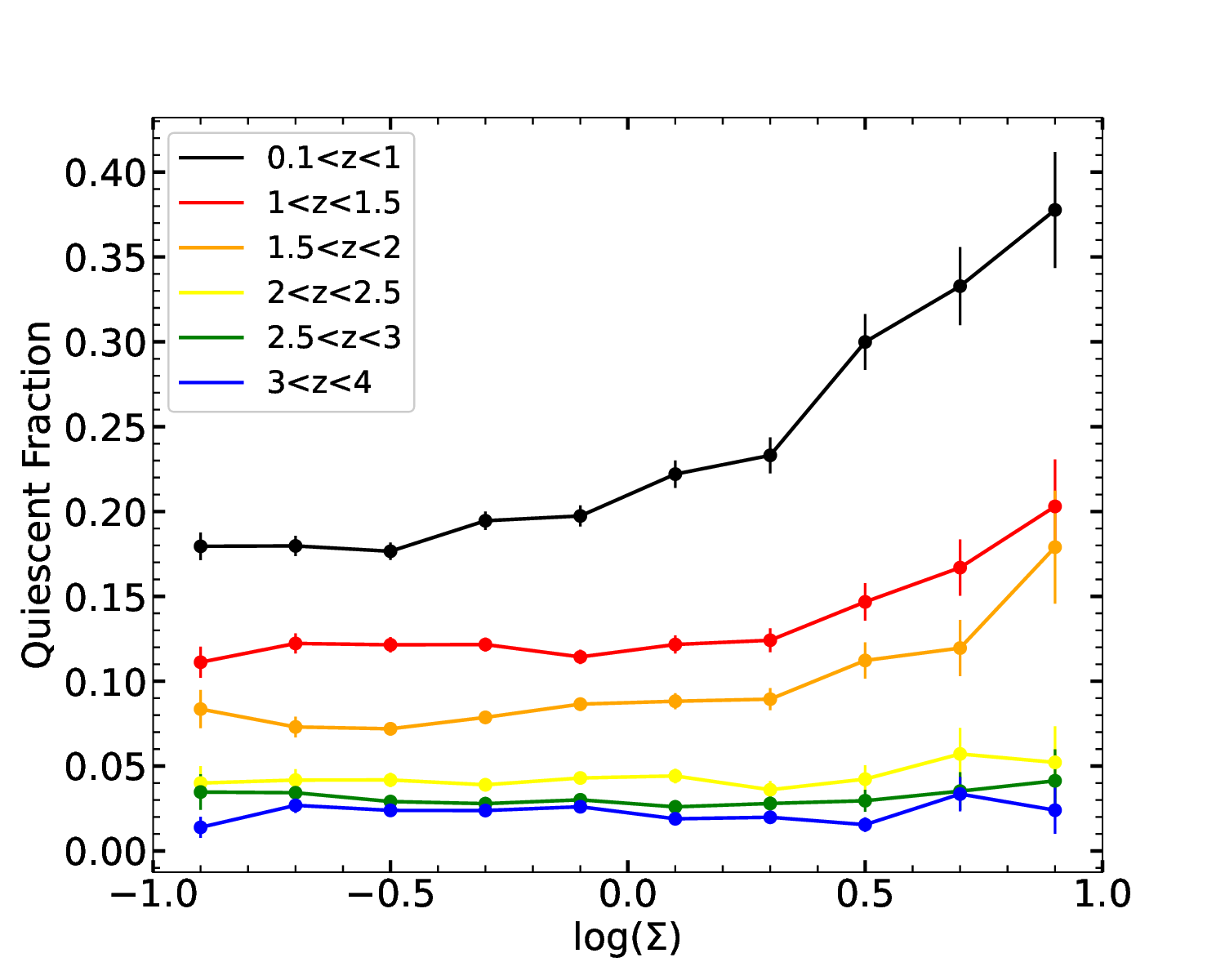}{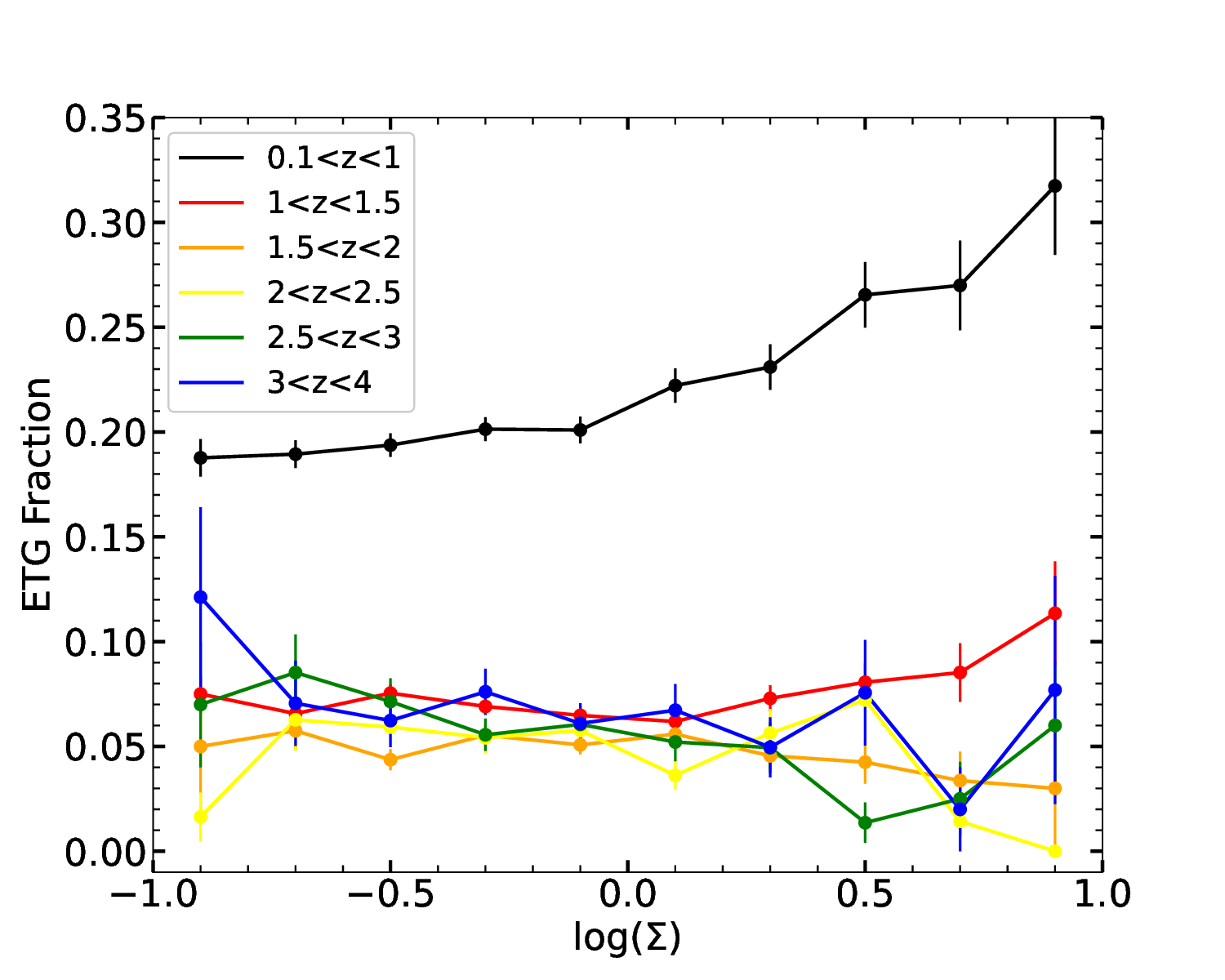}
\caption{
 {\it Left:} Quiescent fraction as a function of density contrast at different redshift bins. The errorbar denotes the Poisson error. 
 {\it Right:} The fraction of early type galaxies as a function of density contrast at different redshift bins. The errorbar denotes the Poisson error. 
}
\label{figure5}
\end{figure*}

Figure \ref{figure5} (left) shows the fraction of quiescent galaxies (defined in Section 2.1) as a function of the environment. It is clear that the quiescent fraction is higher in denser environment, up until $z<2$. This is most significant at lower redshift, for example, we see that quiescent fraction increases dramatically with the environment from $\sim0.18$ to $\sim0.38$ at $z<1$. At higher redshift (e.g., $z>2$) however, no obvious environmental dependence of quiescent fraction can be spotted (see also Table \ref{table2}). This is consistent with the results from Section \ref{sec3}, where we observe a strong decline of star-formation activities at $z<1$ and a flat sSFR-density relation at $z>2$. Our results are fully consistent with \cite{Kawinwanichakij17} who investigated environmental and stellar mass effects in  the FourStar Galaxy Evolution (ZFOURGE) survey at $0.5<z<2$, finding a higher quiescent fraction in denser environments in the redshift range they probed. Similar results have been obtained by \cite{Darvish16} where they found a clear positive correlation between quiescent fraction and environment at $z<2$ but no trend was spotted at $2<z<3$. 

It is also noted that galaxies at lower redshifts tend to have a larger fraction of quiescent galaxies than those at higher redshifts. This is consistent with the ``downsizing'' scenerio \citep{Cowie96, Bundy06, Fontanot09}, in which higher mass galaxies cease their star-formation earlier than lower mass systems. Since structures form in a hierarchical way, galaxies at higher redshifts are usually less massive and the overall star-formation rate density is also higher \citep{Madau14}, resulting in a lower quiescent fraction at high redshifts.

While ``morphology-density'' relation is well established in the local Universe and at low redshift ($z<1$), it is still an open question whether it holds at high redshift ($z>1$). To this end, we take a further step to study the environmental dependence on morphology of the galaxies in the COSMOS field. We make use of Tasca Morphology Catalog (v1.1) \citep{Tasca09} which contains morphological information of 237,192 sources in the COSMOS field that have Hubble Space Telescope (HST) Advanced Camera for Surveys (ACS) observations \citep{Leauthaud07}. The morphological parameters are estimated using the \textsc{Morpheus} code \citep{Abraham07}. To transform morphological parameters into morphological classes, they first eyeballed a training set of $\sim$500 galaxies into early, spiral and irregular types, and performed machine learning technique to derive the types of remaining galaxies with the accociated morphological parameters. We cross-match Tasca catalog with our catalog using 1$\arcsec$ radius, finding 83,779 counterparts.

Figure \ref{figure5} (right) shows the fraction of early type galaxies (ETG) as a function of the environment. We see a significant positive correlation between the ETG fraction and the density of the environment at $z<1$. On the other hand, no obvious environmental trend can be seen beyond $z>1$ (see also Table \ref{table2}). Our work suggests that the ``morphology-density'' relation holds at least till $z\lesssim1$, which is in agreement with many studies in the literature \citep{Dressler97,Goto03,Kauffmann04,Postman05,Smith05,Tasca09}. At low redshift ($z<1$), the evolution of ETG fraction is generally consistent with that of the quiescent fraction. This is expected since early type galaxies usually have much lower SFRs than late type galaxies at low redshift, thus more likely to be quiescent. However, we note that in dense environments (log~$\Sigma>0$), the quiescent fraction seems to be larger than ETG fraction at $z<1$, probably indicating that galaxies quenched their star-formation before the transformation of their morphologies. 

Although we do not find clear environmental effects on galaxy morphology at high redshift ($z>1$) in this work, it should be noted that there have been many studies in recent years confirming the ``morphology-density'' relation at high redshift in cluster environments \citep{Newman14,Sazonova20,Mei23,Strazzullo23}. For example, \cite{Sazonova20} studied the morphology of galaxies in four clusters at $1.2<z<1.8$ with HST imaging, finding two of them have enhanced fraction of bulge-dominated galaxies. \cite{Mei23} further studied 16 clusters at $1.3<z<2.8$ with HST observations, revealing a strong correlation between local environment and ETG fraction. These studies imply that the ``morphology-density'' relation may already be in place in some dense cluster environments at higher redshift, showcasing the early environmental impact on galaxy morphology. The absence of ``morphology-density'' relation at $z>1$ in our work could be due to the fact that COSMOS field does not host a large population of clusters, and our photometrically defined environment could dilute possible signals as discovered in the above spectroscopically confirmed clusters.

\section{Discussion} \label{sec4}
\subsection{On Environmental and Mass Quenching Efficiencies}
In Section \ref{sec3} we showed that the quenching of galaxies could be due to either environment or stellar mass, and it appears that stellar mass might be the dominant mechanism for galaxy quenching (Figure \ref{figure3}). In order to further quantify the efficiency of environment and stellar mass on quenching the star-forming activities in galaxies, we adopt the method introduced in \cite{Peng10} (see also in \cite{Darvish16}, \cite{Kawinwanichakij17} and \cite{Chartab20}) to disentangle the effects of environment and stellar mass. 

We define the environmental quenching efficiency, $\varepsilon_\textrm{env}$, as the fraction of galaxies at a given stellar mass $M$ that would be forming stars in low dense regions, but instead are quenched in high dense regions due to enviromental effects:

\begin{equation}
\varepsilon_\textrm{env}(\Sigma,\Sigma_0,M,z)=\frac{f_q(\Sigma,M,z)-f_q(\Sigma_0,M,z)}{1-f_q(\Sigma_0,M,z)},
\end{equation}

where $f_q$ is the quiescent fraction of galaxies with stellar mass $M$ and density contrast $\Sigma$ at a given redshift $z$. $\Sigma_0$ is the density contrast of the low density environment, which we choose to be the lowest 25 percentage of the $\Sigma$ distribution ($\Sigma_{25}$) as reference following \cite{Kawinwanichakij17}. We calculate environmental quenching efficiency for galaxies that are located in regions with $\Sigma$ greater than  the 75 percentile of the $\Sigma$ distribution. 

Similarly, mass quenching efficiency $\varepsilon_\textrm{mass}$ can be defined as:

\begin{equation}
\varepsilon_\textrm{mass}(\Sigma,M_0,M,z)=\frac{f_q(\Sigma,M,z)-f_q(M_0,\Sigma,z)}{1-f_q(M_0,\Sigma,z)},
\end{equation}

where $M_0$ is the lowest stellar mass at that redshift, which we choose to be in the range of $10^{9.64}M_\odot\sim10^{9.84}M_\odot$ (i.e., larger than 0.2 dex of the mass limit of quiescent galaxies). Following \cite{Kawinwanichakij17}, we calculate the mass quenching efficiency for galaxies with $\Sigma$ smaller than 75 percentile of its distribution. We note our following results do not change if we tweak the values of $M_0$ or $\Sigma$. 

\begin{figure*}[ht!]
\epsscale{1.0}
\plottwo{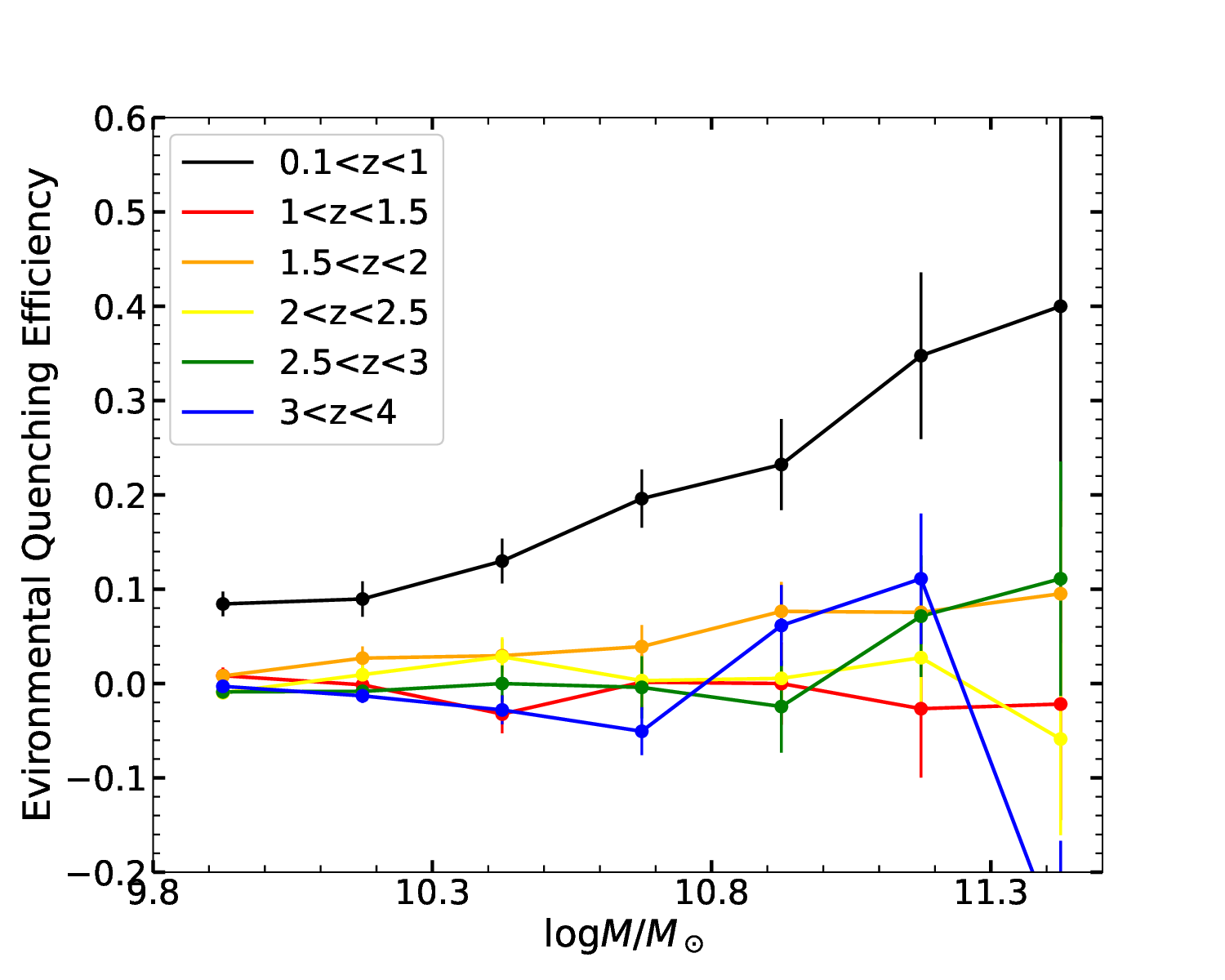}{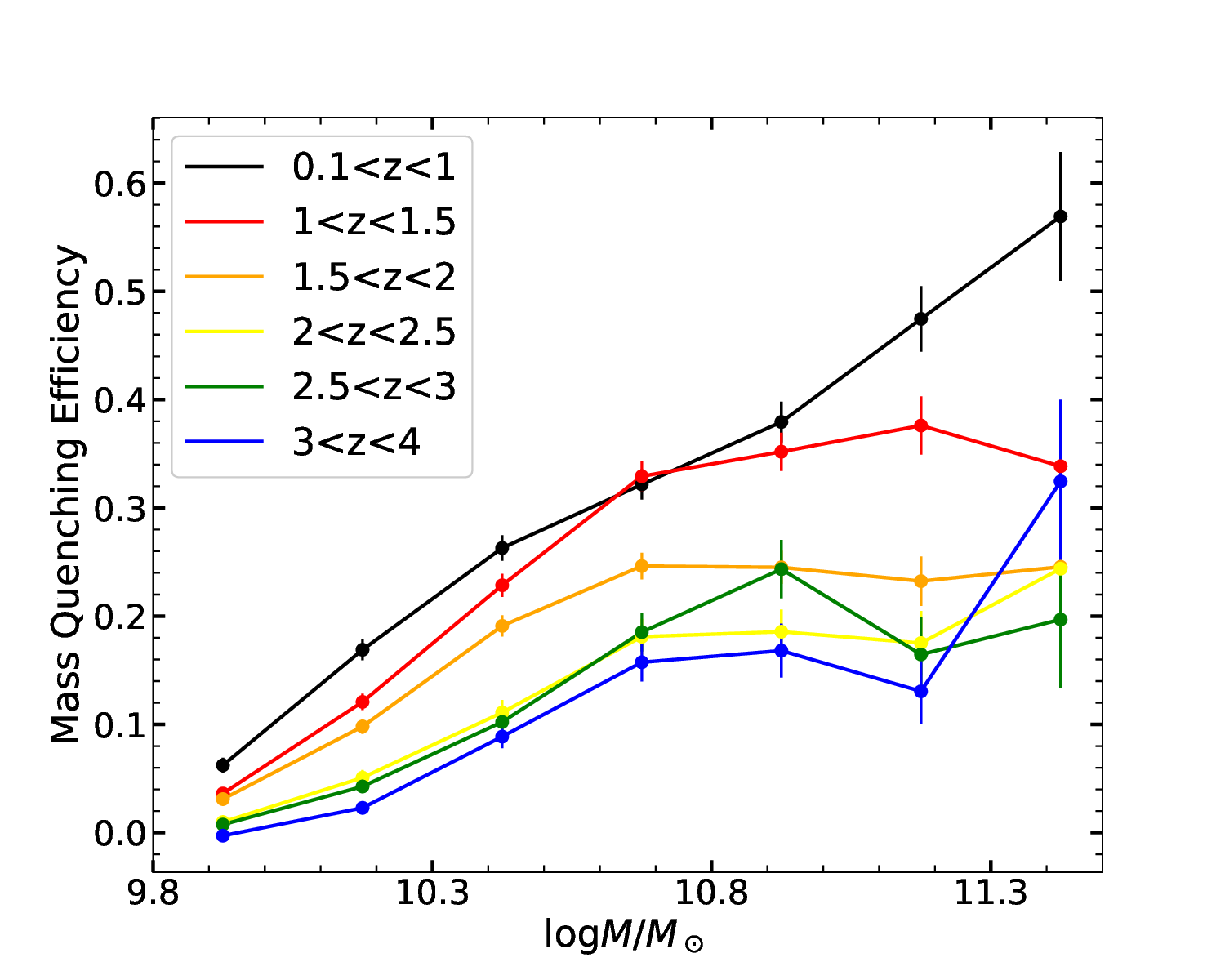}
\caption{
 {\it Left:} The environmental quenching efficiency as a function of stellar mass. The errorbar denotes the Poisson error. 
 {\it Right:} Same as the left, but for mass quenching efficiency.
}
\label{figure6}
\end{figure*}

Figure \ref{figure6} (left panel) shows the environmental quenching efficiency as a function of stellar mass. We find that the environmental quenching efficiency is most prominent at low redshift (e.g., $z<1$), where it increases dramatically with mass from $\sim0.1$ ($10^{9.9}M_\odot$) to $\sim0.4$ ($10^{11.4}M_\odot$).  At $z>1$, the environmental quenching efficiency is almost negligible ($<0.1$). This result is in alignment with that in Section \ref{sec31}, where we see in Figure \ref{figure3} that at $z>1$ the SFRs are almost independent of the environments.

The right panel of Figure \ref{figure6} shows the mass quenching efficiency as a function of stellar mass. Overall, mass quenching efficiency increases with cosmic time, in line with structure formation model. Unlike environmental quenching efficiency which is only relevant at $z<1$, mass quenching efficiency increases significantly with stellar mass at all redshifts. At low redshifts ($z<1$), mass quenching is smaller than environmental quenching for low-mass galaxies (e.g., $M<10^{10}M_\odot$). For instance, environmental quenching is $\sim0.08$ while mass quenching is $\sim0.06$ for galaxies of $M=10^{9.9}M_\odot$. While for massive galaxies ($M\gtrsim10^{10}M_\odot$) mass quenching begin to dominate over environmental quenching. For example, at $z<1$, environmental quenching efficiency is $\sim0.3$ while mass quenching efficiency is $\sim0.4$ at $10^{11}M_\odot$. On the other hand, at $z>1$ environmental quenching is $\lesssim0.1$ for all galaxies  while mass quenching is already  $>0.1$ for galaxies with mass $>10^{10.5}M_\odot$ even in the highest redshift bin.

Our results suggest that mass quenching is the dominant quenching mechanism for massive galaxies ($M\gtrsim10^{10}M_\odot$) at all redshifts, whereas environmental quenching only takes effect at low redshift ($z<1$) and only exceeds mass quenching for less massive galaxies (e.g., $M\lesssim10^{10}M_\odot$). The results in this section reinforce our arguments in Section \ref{sec31}, that massive galaxies are primarily quenched by their mass rather than the environments at low redshifts. In this sense, the different ``SFR-density'' relations we observed at different redshifts could largely be attributed to the stellar mass.   

Our result is in good agreement with \cite{Darvish16}, who also studied environmental effects in the COSMOS field using an early version of the COSMOS catalog. Due to the limited depth of their data, they only considered galaxies with $z<3$ (73,481 in total), finding that environmental quenching is only prominant at low redshift ($z<1$) for massive galaxies whereas mass quenching is the dominant quenching mechanism at $z>1$. Similar results have been obtained by \cite{Kawinwanichakij17} who showed that environmental quenching efficiency is minimal for low mass ($M\lesssim10^{10}M_\odot$) galaxies at high redshift ($z>1.5$), but increases rapidly with decreasing redshift, till it dominates over mass quenching for those less massive galaxies at $z<1$.

Lastly, as shown in Figure \ref{figure6} (left), environmental quenching does depend on stellar mass for massive galaxies at low redshift, which was also noticed in \citet{Darvish16,Kawinwanichakij17,Chartab20}. The two processes (internal and external) may actually be dependent on each other, thus it would be difficult to completely disentangle mass and environmental quenching.

\subsection{The Possible Mechanisms of Galaxy Quenching}

The physical processes that are responsible for galaxy quenching are still not well understood and in hot debate. Here we discuss several scenerios that may in part explain the cause of  quenching we observed in this work. 

In Figure \ref{figure6} we see that at high redshift ($z>1$), environmental quenching is almost negligible while mass quenching becomes prominent with increasing mass. \cite{McGee14} proposed that internal outflows may play a major role in removing the gas from satellite galaxies in host halos.  In this ``over-consumption'' model, satellite galaxies cease their growth by outflows produced from vigorous star-formation or AGN feedback, which rapidly exhaust the remaining gas reservoir (``starvation''). The gas depletion time $\tau_\textrm{dep}$ is proportional to the ratio of gas mass to star-formation rate ($\tau_\textrm{dep}\propto M_\textrm{gas}/SFR$). Since more massive galaxies usually have higher star-formation rate, given the same amount of gas, $\tau_\textrm{dep}$ becomes shorter for more massive galaxies. In addition, this model also suggests that the quenching timescale is shorter ($<200$ Myrs) at high redshift ($z>1.5$) because of higher star formation rates \citep{McGee14,Balogh16}. Therefore this ``over-consumption'' model could partly explain the quenching of massive galaxies at high redshift ($z>1$). However, this mechanism cannot fully account for the quenching of satellites at very low redshift ($z\ll1$). At these redshifts, as star formation rates and associated outflow rates decline, \cite{McGee14} predicted that the quenching time will reach $\sim$10 Gyrs, which is much longer than the dynamical time \citep{Balogh16}. Therefore other internal processes such as AGN or stellar feedback may be responsible for quenching of low redshift galaxies.

As for environmental quenching, it is often thought that hydrodynamic process like ram pressure stripping is one likely candidate mechanism for the cause of quenching in dense environments such as galaxy clusters \citep[e.g.,][]{Vollmer01,Boselli06,Boselli22}.
What is more, many studies have found that ram pressure stripping becomes increasingly effective for low-mass satellite galaxies ($M\lesssim10^{8}M_\odot$) \citep[e.g.,][]{Slater14,Fillingham16,Simpson18,Boselli22}. Athough we can only probe galaxies with $M\gtrsim10^{10}M_\odot$, the general trend in Figure \ref{figure6} implies that environmental quenching would be more effective than mass quenching for low-mass galaxies ($M\lesssim10^{10}M_\odot$). Therefore ram pressure striping might play a major role in quenching of low-mass galaxies.  Aside from ram pressure striping, major mergers are also considered to be an important quenching mechanism \citep[e.g.,][]{Springel05,Gabor10,Treister12,Pontzen17}. As two gas-rich galaxies merge (which are expected to be more frequent in dense environments), the inflow of cold gas often trigers a starburst or an AGN, which can rapidly exhaust or heat the gas, and in turn quench the remnant galaxy. 
 
We conclude that ``over-consumption'' model is likely an effective internal quenching mechanism for massive galaxies at high redshift observed in this work. At low redshift however, external processes such as ram pressure stripping and galaxy mergers might become more efficient in halting the star formation of less massive galaxies.

\subsection{Reversal of SFR-density Relation at High Redshift?}
In Section \ref{sec31}, we find clear evidence that the ``SFR-density'' relation observed in local Universe is reversed at high redshift ($z>2$), in a way that galaxies in denser environments tend to have higher star-formation rates. However, this reversal of ``SFR-density'' relation at high redshift is still controversial and different studies often yield different results. In this section we give a tentative comparison between different studies and discuss possible causes for the reversal discovered in this work.

Using an earlier version of the COSMOS catalog, \cite{Darvish16} found that SFR and sSFR decrease with increaing density of the environment at $z\lesssim1$, but become almost independent of the environment at $z\gtrsim1$. They argued that denser environments increase the fraction of quiescent galaxies at low redshift, leading to the ``SFR-density'' relation observed at $z\lesssim1$. The major difference between their work and ours is the ``SFR-density'' relation at high redshift,  where we find a significant positive correlation at $z\gtrsim2$. We argue that since denser environments generally host more massive galaxies, the higher SFRs of these massive galaxies can naturally result in the reversal of ``SFR-density'' relation at high redshifts. On the other hand, at low redshift ($z\lesssim1$) due to some mass quenching mechanisms, SFRs decline dramatically for those massive galaxies, leading to the observed ``SFR-density'' relation in the local Universe. 

It should be noted that due to the limited depth of their data, the highest redshift sample ($2<z<3$) in \cite{Darvish16} only reached a mass completeness of $\sim10^{10}M_\odot$. In comparison, our sample reaches $\sim10^{9.3}M_\odot$ at that redshift. Thus the lack of low-mass galaxies in their sample could introduce a selection bias which affects the study of environmental trend. In Figure \ref{figure6} we see that mass quenching is non-negligible for massive high redshift galaxies (from $\sim0.1$ at $10^{10.3}M_\odot$ to $\sim0.2$ at $10^{10.8}M_\odot$), as a result the massive galaxies in the highest redshift bin of \cite{Darvish16} could severely suffer from mass quenching. Since they only used these massive galaxies to define the local environments, and more massive galaxies (more likely to be quiescent) prefer higher density regions, therefore they observed a flat ``SFR-density'' relation at high redshift.
Alternatively, the larger photometric errors in their catalog could also dilute the possible signal. For instance, as we noted in Section \ref{data}, many high redshift galaxies selected in the COSMOS2020 catalog are actually found to be at low redshift with far-infrared to sub-mm data \citep{Jin18}. The data \cite{Darvish16} utilized could suffer even more from this contamination due to larger photometric redshift uncertainties in the early version of COSMOS catalog \citep{McCracken12,Ilbert13}. As a consequence, these low redshift galaxies can possibly bias the the environmental trend by their lower SFRs, leading to a flat ``SFR-density'' relation in their highest redshift bin.

\cite{Chartab20} explored the environmental effects of a mass-complete sample of $\sim20$k galaxies in the CANDELS field out to $z\sim3.5$, finding that the average SFR and sSFR both decrease with increasing density of the environments at all redshifts. Same issues as in \cite{Darvish16}, we note their sample only reached a mass completeness of $10^{10.3}M_\odot$ at $2.2<z<3.5$. Therefore their sample could suffer even more from mass quenching than \cite{Darvish16}, leading to a normal ``SFR-density'' relation as in the local Universe.

The reversal of ``SFR-density'' relation at high redshift observed in our study is fully consistent with the work done by \cite{Lemaux22}, who found clear evidence of a monotonic increase in the SFR with increasing galaxy overdensity at $2<z<5$ using spectroscopic observations from the VIMOS Ultra-Deep Survey (VUDS). They also found that stellar mass increases with increasing galaxy overdensity, in agreement with our study. They argued that the reversal of ``SFR-density'' relation observed at high redshift is primarily driven by the fractional increase of massive galaxies in dense environments. Similar conclusion has been drawn from \cite{Momose22}, who noted that stellar mass might play a major role in the star-formation activities of galaxies in their reconstructed 3D density fields from galaxy and Ly$\alpha$ forest spectroscopy at $2<z<2.5$ in the COSMOS survey. 

Last but not least, the controversy in the ``SFR-density'' relation in the literature could be due to selection effects, such as the dynamical state of the environment in the study. This is particularly the case in cluster or protocluster study, such that cluster environments automatically bias towards a normal ``SFR-density'' relation observed in the local Universe, while the reversal of SFR-density relation is often found in protocluster environments \citep[e.g.,][]{Tran10,Shimakawa18,Shi20} due to enhanced star-formation activities therein.

To conclude, although it is still an open question about the reversal of ``SFR-density'' relation at high redshift, we argue that the discrepancies of different studies could be due to selection bias as well as the different dynamical states of the environments probed. Future wide-field surveys such as 
LSST \citep{LSST} and Euclid \citep{Euclid} could shed light on this topic and improve our understanding of the environmental impacts on galaxy formation.

\section{Summary and Conclusion} \label{sum}
In this paper, we probe the environmental effects on galaxy formation by utilizing the latest COSMOS2020 photometric catalog. After defining the environments by the Voronoi tessellation method, a mass-complete sample of 173,339 galaxies at $0.1<z<4$ is selected to study the role of environments in the star-formation activities of galaxies. Our major results are summarized below.

1. At $z<1$, galaxies exhibit decreasing SFR at increasing density of the environments in a statistic manner, consistent with the normal ``SFR-density'' relation observed in the local Universe. When consider only star-forming galaxies, their mean SFRs become independent of the environment, indicating that dense environments increase the fraction of quiescent galaxies within.  At $1<z<2$ the mean SFRs become independent of the environments for all the galaxies. However at $z>2$, there is a clear positive correlation between SFRs and environments. At almost all redshifts, the mean stellar mass of all the galaxies increases monotonically with the environments. The only exception is star-forming galaxies at $z<1$, where no obvious correlation has been found, suggesting that denser environments tend to host massive quiescent galaxies at low redshift.  

2. At $z<2$, we find that the fraction of quiescent galaxies increases with increasing density of the environments, indicating that galaxies quench more rapidly in denser environments. Meanwhile, we show that the fraction of early type galaxies increases with increasing density of the environments at $z<1$, confirming the presence of ``morphology-density'' relation at least up to $z\sim1$.

3. We find that environmental quenching is only relevant at low redshift ($z<1$) but negligible at high redshift ($z>1$). On the other hand, mass quenching is prominent for galaxies at all redshifts, which increases significantly with stellar mass. At low redshift ($z<1$), mass quenching dominates over environmental quenching for massive galaxies ($>10^{10}M_\odot$).

4. The quenching of massive galaxies at high redshift could be partly explained by the ``overconsumption'' model, in which the gas reservoir is rapidly consumed by vigorous star formation. At low redshift however, dynamical processes such as ram pressure striping or mergers could partly be responsible for the quenching of less massive galaxies.

5. We argue that stellar mass might be the major driving force for star formation activities of galaxies. In general, galaxies in denser environments are more massive. At high redshift ($z>2$), the sample is dominated by star-forming galaxies, among which more massive ones have higher SFRs, naturally resulting in a reversal of ``SFR-density'' relation. While at low redshift ($z<1$), massive galaxies are drastically quenched due to some internal processes (i.e., mass quenching), which leads to declined SFRs with the environments.
The discrepancy of the ``SFR-density'' relation observed in different studies might be attributed to sample bias or different dynamical states of the environments probed.

We thank the anonymous referee for a careful review and helpful comments that improved this paper. K.S. thank Shuowen Jin and Chong Ge for helpful discussions. K.S. acknowledge the funding from NSFC grants No. 12003023 and Fundamental Research Funds for the Central Universities under Grant No. SWU-KR22035.

\bibliography{swu}
\bibliographystyle{aasjournal}



\end{document}